\documentclass{emulateapj}

\usepackage{graphicx}
\usepackage{apjfonts}
\usepackage{mathptmx}



\def\***#1{{\sc #1}}
\def\plan#1{\relax}
\def\Plan#1{\relax}
\def\PLAN#1{\relax}

\def\lta{\mathrel{\spose{\lower 3pt\hbox{$\mathchar"218$}}
     \raise 2.0pt\hbox{$\mathchar"13C$}}}
\def\gta{\mathrel{\spose{\lower 3pt\hbox{$\mathchar"218$}}
     \raise 2.0pt\hbox{$\mathchar"13E$}}}
\newcommand{\etal}{{\it et al.}}

\shorttitle{Non-linear behavior of GRS 1915+105}
\shortauthors{Misra \etal}

\def\mathnew{\mathsurround=0pt}

\def\simov#1#2{\lower .5pt\vbox{\baselineskip0pt \lineskip-.5pt
\ialign{$\mathnew#1\hfil##\hfil$\crcr#2\crcr\sim\crcr}}}

\begin{document}

\title{The non-linear behavior of the black hole system GRS 1915+105}

\author{R. Misra\altaffilmark{1}, K. P. Harikrishnan\altaffilmark{2},
G. Ambika\altaffilmark{3} and A. K. Kembhavi \altaffilmark{1}}

\altaffiltext{1}{Inter-University Center for Astronomy and Astrophysics, Post Bag 4,
Ganeshkhind, Pune-411007, India email: rmisra@iucaa.ernet.in}
\altaffiltext{2}{Dept. of Physics, The Cochin College,  Cochin-682002, India}

\altaffiltext{3}{Dept. of Physics, Maharajas College,  Cochin-682011, India}

\begin{abstract}
Using non-linear time series analysis, along with surrogate data
analysis, it is shown that the various types of long term variability
exhibited by the black hole system GRS 1915+105, 
can be explained in terms of a deterministic non-linear system with
some inherent stochastic noise.
Evidence is provided for a non-linear limit cycle origin of one of 
the low frequency QPO detected in the source, while some other
types of variability could be due to an underlying low dimensional chaotic
system. These results imply that the partial differential equations which govern
the magneto-hydrodynamic flow of the inner accretion disk, can be approximated
by a  small number ( $\approx 3 -5$) of
non-linear but {\it ordinary} differential equations.
While this analysis does not reveal
the exact nature of these approximate equations, they may be obtained in the future, 
after results of magneto-hydrodynamic simulation of realistic accretion 
disks become available.

\end{abstract}


\section{Introduction}

Black hole X-ray binaries exhibit variability 
on a wide range of timescales ranging from months to milli-seconds.
This variability is expected to provide important clues about
the geometry and
structure of these high energy sources. Moreover, a detailed
study of the phenomena may eventually 
be used to test 
the relativistic nature of these sources and to understand the physics
of the accretion process.

The standard first step to quantify variability is to compute
the power spectrum which is the amplitude squared of the Fourier transform
of the light curve. The power spectra is expected to  give
information about characteristic frequencies of the system which might
show up  either as spectral breaks or as near Gaussian peaks, i.e Quasi-Periodic
Oscillations (QPO) \citep[e.g.][]{Bel01,Tom01,Rod02}.
The shape of the power spectra, combined with the
observed frequency dependent time lags between different energy bands,
have put constraints on the radiative mechanism and geometry of the
emitting region  \citep[e.g.][]{Now99,Mis00,Cui99,Pou99,Nob01}.

The accretion disk which powers these sources maybe undergoing
stochastic variations at different length and time scales. 
This is expected, especially since the viscosity mechanism which
drives the accretion process is now believed to be caused by turbulence
induced by magnetic instabilities. These structural variations
may then manifest as variability in the observed intensity after
being convolved with the main radiative process that cools the disk.
Additionally, these local stochastic variations could couple 
resonantly with the global disk structure and produce coherent 
oscillations. In most analysis, it is implicitly assumed that
response of the disk to these fluctuations, either by global
resonance and/or radiative processes, is linear in nature.  
However, there is some evidence that the 
response of the disk may be non-linear.
The ratio of the twin high frequency QPO observed in black hole and
neutron star systems, can be explained in a model where the
global disk resonates nonlinearly with the stochastic fluctuations
\citep{Aba03}. More compelling model independent evidence has been
given by \cite{Utt05} (see also \cite{Tim00} and \cite{Thi01}), 
who argue that the log-normal distribution
of the fluxes and the linear relationship between RMS and flux
imply that the response of the disk is non-linear. They show
that an exponential response can explain these observations.
Earlier \cite{Min94}, had suggested that the behavior of
the observed fluctuations could be because the accretion disk
is in self-organized critical state. 
In all these models the  temporal behavior of the system is
driven by underlying stochastic variations. Moreover, 
they address the short ($ < 10 $ sec) time-scale  variability 
of the systems, although many systems also exhibit quite dramatic
long time-scale variability.

Variability of a system may not necessarily be driven by an
underlying stochastic variation. The system can show complicated
temporal behavior if the governing differential equations are
non-linear and have unstable steady state solutions. In other
words, although these systems do not have an explicit time
dependent term in the equations describing their structure,
they exhibit sustained time variability. In fact, the
standard accretion disk theory predicts that the disk 
is unstable when it is radiation pressure dominated and
when the viscous stress scales with the total pressure.
Numerical hydro-dynamic simulations reveal that under
such circumstances, the disk would undergo large amplitude
oscillations around the unstable solution \citep{Che94}. 
These variations occur on a viscous time-scale that
may be as large as hundreds of seconds. Specific dynamic
models  for the temporal behavior of accretion disk, 
like the Dripping Handrail \citep{You96}, have been proposed
where the apparent random behavior is actually deterministic.

The Galactic micro quasar GRS 1915+105 is a highly variable black
hole system. It shows a wide range of long term 
variability \citep{Che97,Pau97,Bel97a} 
which required
\cite{Bel00} to classify the behavior in no less than twelve temporal classes.
For some classes, low frequency oscillations are observed and the
light curves resemble those from numerical simulations of unstable
radiation pressure dominated disks based on the standard accretion
disk theory. However, the viscosity prescription in the standard
disk theory is ad hoc, and while it may represent approximately the time
averages viscosity, it is not expected to reproduce the complex
time-varying turbulent induced viscosity expected in an accretion disk.
Moreover, there could be different types of instabilities in the disk which
could give rise to strong variability.
Nevertheless, the qualitative similarities between the results of
the simulation and observations \citep{Taa97}, 
suggests that the long term variability of this source is due to 
deterministic non-linear evolution of an unstable disk, rather than
a manifestation of an underlying stochastic process. 

It is 
important to determine in a more quantitative and general method,
whether the long term variability is indeed due to deterministic
non-linear behavior rather than having a stochastic origin. That
would imply that the temporal behavior of the system can be described
by a small number of non-linear ordinary differential equations. In other
words, the complex non-linear partial differential equations that are
known to govern the hydrodynamical flow, can be approximated by a set
of ordinary differential equations and hence can be more easily studied
and understood. Such an approximation can be obtained, for example,
by studying the non-linear time evolution of a dominant linear mode. This technique
which is briefly described in the next section, was used to derive the equations for
the well known non-linear Lorenz system. A more physical method is to
study the time evolution of  spatially averaged quantities as in
the one zone approximation by \citet{Pac83} for thermo-nuclear flashes
on compact objects. While the rewards of obtaining such approximated equations
are far-reaching, these methods are in general not straight forward and
require good physical intuition, especially since it is not necessary that
they can be obtained for all systems.  Hence, evidence for the existence
of such an approximation, would give the necessary impetus and direction
to the study of these complex systems.  

Non-linear time series analysis  have been used earlier to
analyze  X-ray light curves of astrophysical sources. 
\cite{Leh93}
used the  correlation dimension technique to analyze  
EXOSAT light curves of several AGN,  and found that one, NGC 4051,
showed signs of low dimensional chaos. Such analysis
have been undertaken on noise filtered {\it Tenma} satellite data of Cyg X-1
\citep{Unn90} and on EXOSAT data of Her X-1 \citep{Vog87,Nor89} and
ASCA data of AGN ARK 567 \citep{Gli02}.
In an earlier work, we used a modification of this technique
to study RXTE observations of GRS 1915+105 and found evidence
that for four of the twelve temporal classes, the system was
consistent with harboring a low-dimensional attractor \citep{Mis04}. 

However, a positive result for the correlation dimension analysis,
although highly suggestive, is not conclusive evidence that the
light curve of these systems is due to a low dimensional chaotic
system. The results need to be counter checked using other non-linear
techniques and compared with surrogate data analysis. Thus, in
this work, we analyze the different temporal patterns exhibited
by GRS 1915+105 using  correlation dimension analysis and
the singular value decomposition technique. For each analysis,
we undertake surrogate data analysis as a counter check. To facilitate
 better physical understanding, we draw analogies between the
results obtained for GRS 1915+105 and those for a 
well known deterministic non-linear system.

It should be emphasized that proving
the presence of non-linear dynamics using a finite time series 
with mathematical certainty is in the very least    
difficult and perhaps even not a well defined problem.
Our motivation here, is not to make model independent
and quantitative non-linear diagnosis of the light curve, which as
mentioned above is difficult and ambiguous. The goal
here is to find out if there are features in the
light curve which indicate (but may not necessarily rigorously prove)
that the temporal behavior of the black hole system 
GRS 1915+105, can be described as a set of ordinary
non-linear differential equations. 
Obtaining such equations
from first principles would be a major break through in our
understanding of accretion disks around black holes and thus
any indication that such equations may exist would be useful
in formulating them.

In the next section, the well known non-linear system, ``Lorenz'', is 
introduced and some of the different types of non-linear behavior that
the system can exhibit are described. In \S 3, some standard 
non-linear time series analysis is described and as an example
applied to the Lorenz system. In \S4, these techniques are applied
to the observed light curves of GRS 1915+105 and in \S5 the results are
summarized and  discussed.

\section{The Lorenz System}

\begin{figure}
\begin{center}
\includegraphics*[width=4cm,height=6cm]{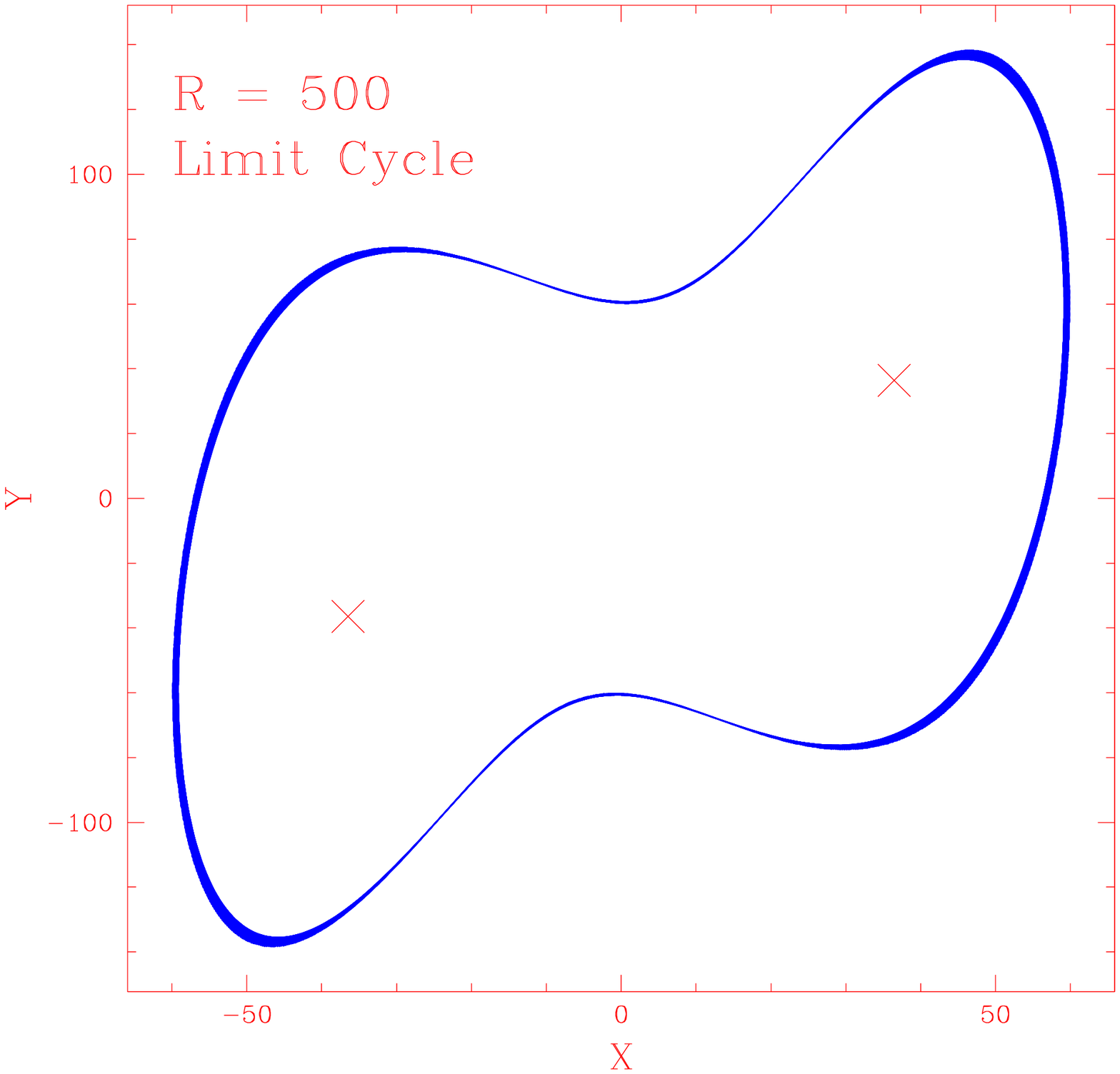}
\includegraphics*[width=4cm,height=6cm]{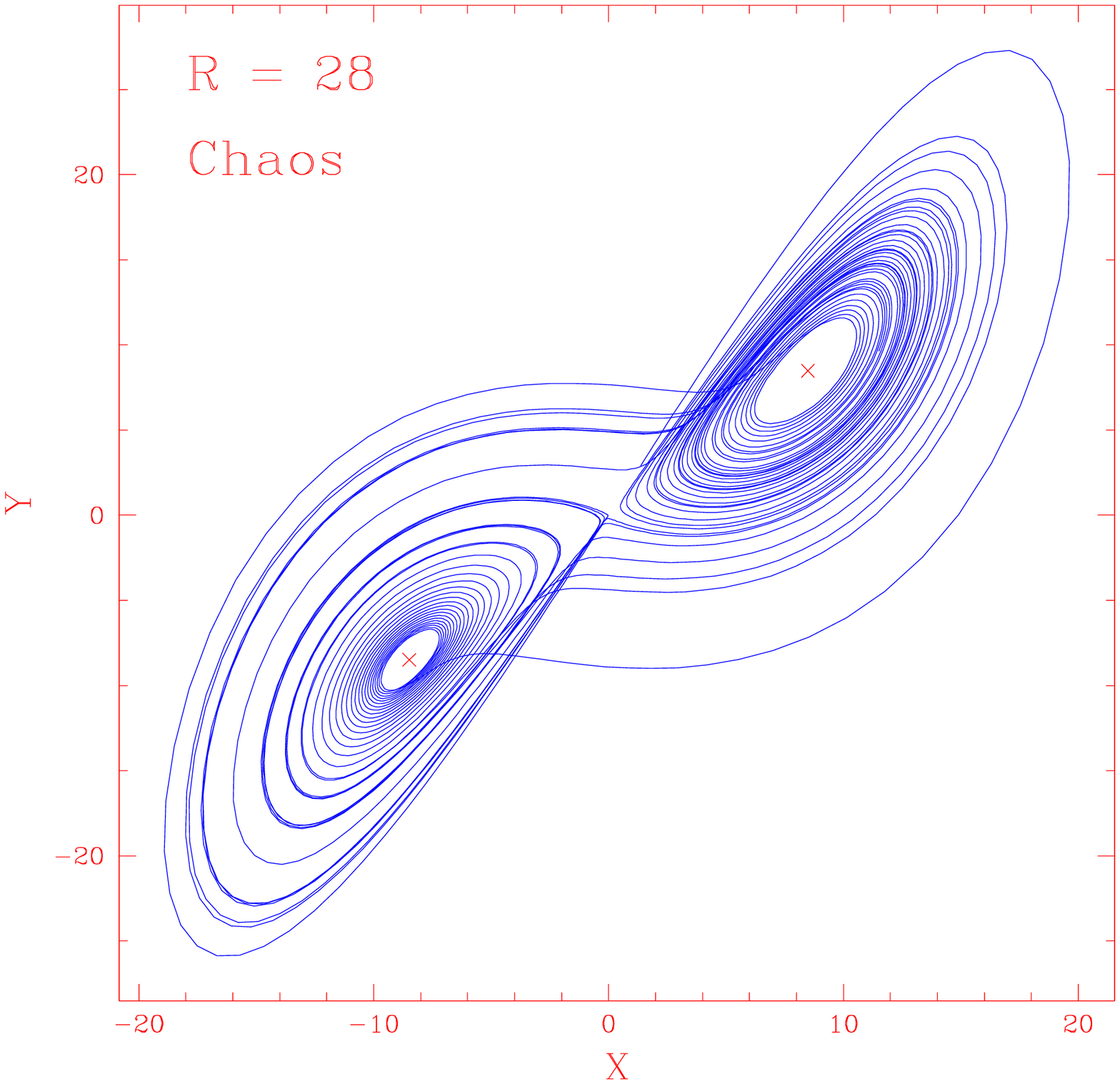}
\end{center}
\caption{ The Dynamics of the Lorenz system. Left Panel is for
$R = 500$ showing a limit cycle while the right panel is for $R = 28$,
when the system exhibits chaos.   }
\label{lormap}
\end{figure}

The Lorenz model \citep{Lor63} was developed from the Navier-Stokes hydrodynamic
equations for the Rayleigh-B\'{e}nard flow, which describes
the two-dimensional convection of an incompressible fluid in a cell which
has a higher temperature at bottom and a lower temperature at top. The essential
idea is to choose a dominant linear mode that satisfies the boundary condition,
and substitute this mode back into the hydrodynamic equations to obtain the
temporal evolution of the mode as a set of ordinary non-linear differential equations.
The choice of the dominant mode is rather arbitrary and hence makes this procedure
possible only by physical intuition and/or prior knowledge of the solution from
experiments or numerical simulations. In spite of the physical (and in particular)
hydrodynamical origin of the Lorenz model, it is prudent to be careful about
drawing analogies between it and other complex hydrodynamical flows
like accretion. For this work, 
it is sufficient to state that the Lorenz model is a mathematical set
of non-linear equations which have been derived from
non-linear partial differential equations, using approximations
which may or may not be valid. These equations turn out to be

\begin{eqnarray}
\dot X & = & 10(X-Y) \nonumber \\
\dot Y & = & R X - Y - XZ \nonumber \\
\dot Z & = & (8/3) Z  - XY 
\label{loreqn}
\end{eqnarray}
There are three  fixed points, or steady state values, of the system:
$ (X_o,Y_o,Z_o) = (0,0,0)$ and $( \pm \sqrt{[8/3][R-1]}, \pm \sqrt{[8/3][R-1]}, [R-1])$.
$R$ is a control parameter whose value governs the system's behavior
which is obtained by solving numerically these differential equations.
For $1 < R < 14$, the fixed point at the origin is unstable, while
the other two are stable. The system  evolves towards one
of the stable fixed points depending on the initial conditions.
The temporal behavior is transitory in nature. For $R > 14$,
all three fixed points are unstable. For large values of $R > 100$,
the trajectory along  $X$ and $Y$ is a closed loop, called
the limit cycle (Left panel of Figure \ref{lormap}). 
For any initial condition the
system evolves towards the limit cycle. For intermediate
values of $R$ the behavior of the system is more complex. The
right panel of 
Figure \ref{lormap} shows the trajectory of $X$ and $Y$ for $R = 28$,
where the behavior is termed chaotic, i.e. infinitesimally close trajectories diverge exponentially in time.  
Thus, the Lorenz system can exhibit
different types of non-linear behavior.

\begin{figure}
\begin{center}
\includegraphics*[width=8cm]{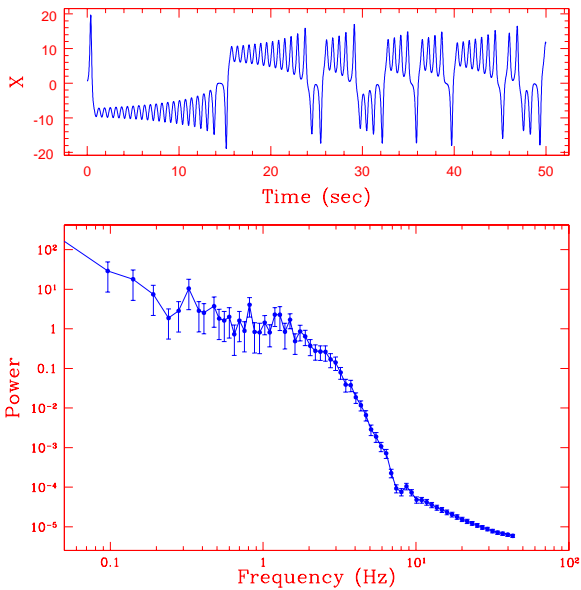}
\end{center}
\caption{ Top panel: The time series $X (t)$ for the Lorenz system
with $R = 28$. Bottom Panel: The corresponding power spectrum.  }
\label{lor_pow}
\end{figure}

Often, in a natural system, only one of the several parameters
can be detected and hence the complete dynamic information
is not directly available. For example, if only the $X$ variable of
the Lorenz system was known, then from the time series for $R = 28$
(top panel of Figure \ref{lor_pow}) 
and the corresponding power spectrum (bottom panel of Figure \ref{lor_pow}), 
the
rich complexity of the Lorenz system is not easily identifiable. 
However, as discussed in the next section, there are
non linear techniques by which the complete (qualitative if
not quantitative) dynamic nature of such systems can be
reconstructed.

\section{Non-linear Time Series Analysis}

\begin{figure}
\begin{center}
\includegraphics*[width=4cm,height=6cm]{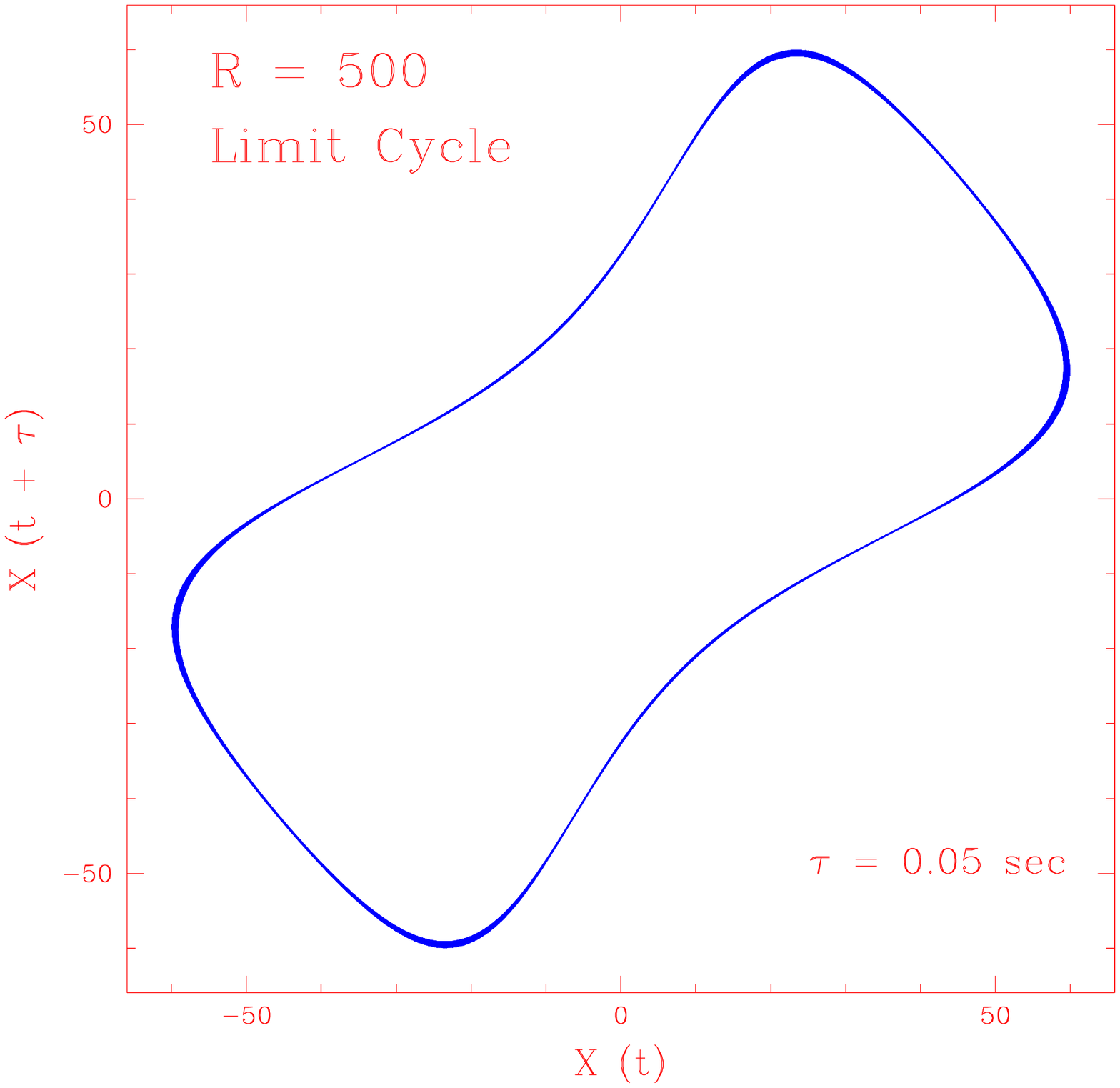}
\includegraphics*[width=4cm,height=6cm]{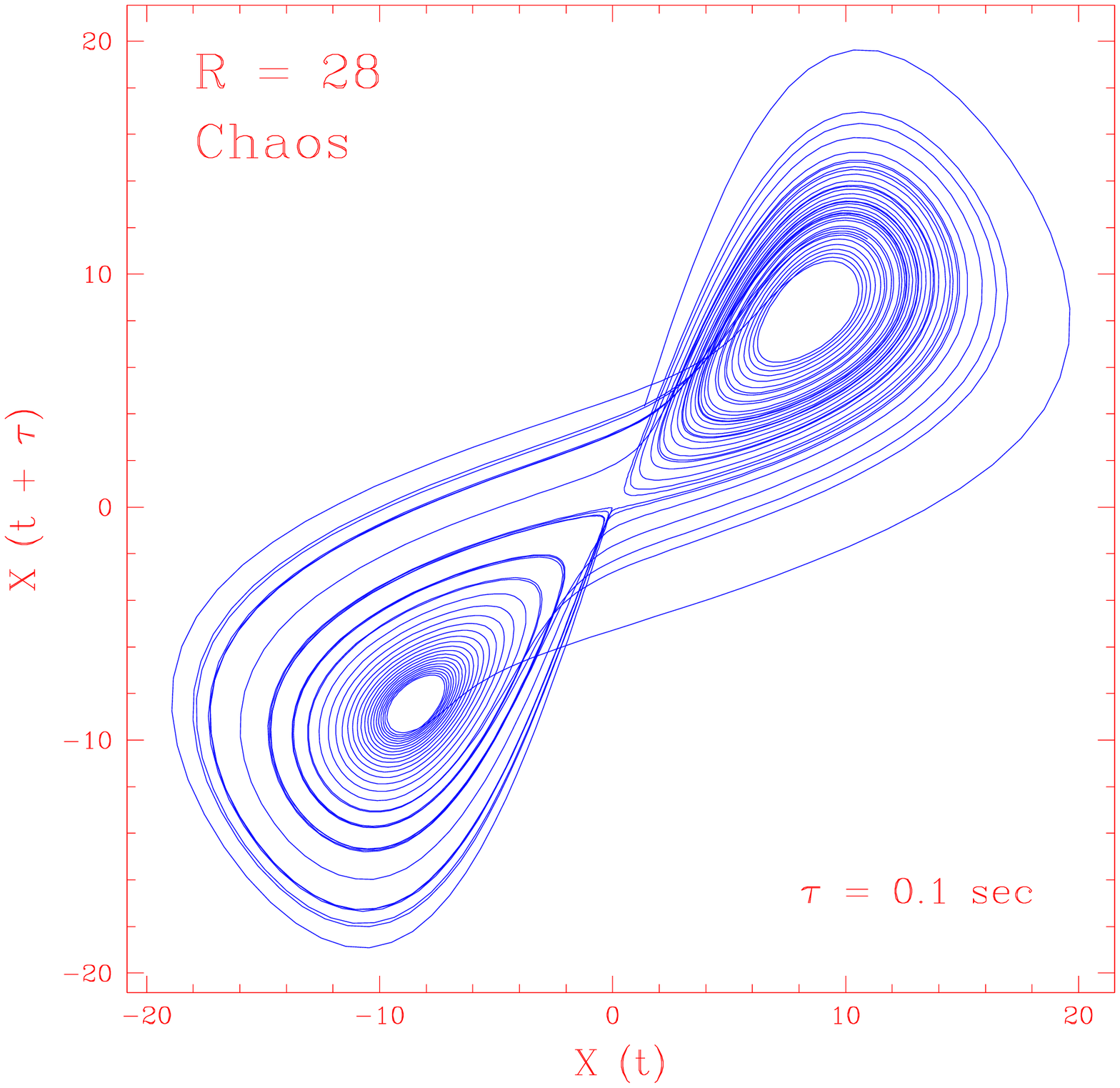}
\end{center}
\caption{ The reconstructed dynamics of the Lorenz system using
the delay embedding scheme. Left Panel is for $R = 500$ while the
right panel is for $R = 28$.  }
\label{lor_Delay}
\end{figure}

One of the standard methods to reconstruct the dynamics of
a non-linear system from a time series, is 
the Delay Embedding Technique \citep{Gra83}. 
Since the dimension of the system ( i.e. the number of
governing equations or variables) is not a priori known,
one has to construct the dynamics for different dimensions.
In this technique, vectors of length $N$ are created from the time series, $X(t)$,
by using a delay time $\tau$, i.e. 
\begin{eqnarray}
 \vec v_1 & = & \;\; (X(1),X(2)...X(N)) \nonumber \\
\vec v_2  & = & \;\;(X(1+\tau),X(2+\tau)...X(N+\tau)) \nonumber\\
       &  &   ... \nonumber \\
\vec v_M & = & \;\; X(1+(M-1)\tau),...X(N+(M-1)\tau)
\end{eqnarray}
where $M$ is the chosen  dimension. 
Typically, the delay time $\tau$ is suitably chosen such
that the vectors are not correlated i.e. when the auto-correlation
function of $X(t)$ approaches zero. For the Lorenz system, such
a technique can effectively reconstruct the dynamics as shown
in Figure \ref{lor_Delay}, where the $X(t)$ is plotted versus
$X(t+\tau)$. Note the similarity between Figures \ref{lor_Delay}
and \ref{lormap}, with the identification of $X(t +\tau)$ with
$Y$.  

\begin{figure}
\begin{center}
\includegraphics*[width=8cm]{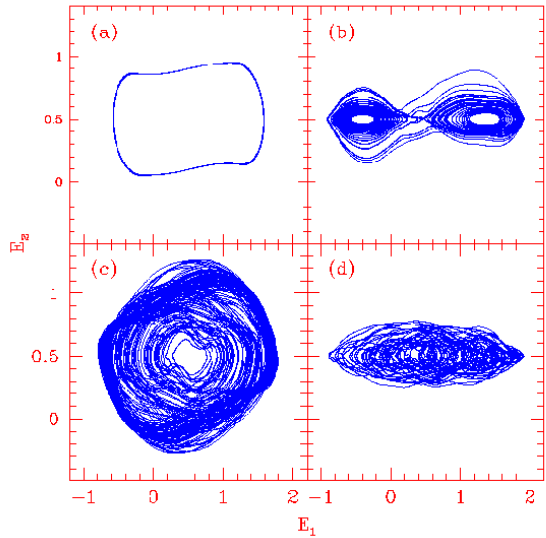}
\end{center}
\caption{ Singular Value Decomposition (SVD) technique
used to reconstruct the Lorenz system dynamics from $X (t)$.
(a) Using real data for $R = 500$, (b)  using real data for $R = 28$,
(c) using surrogate data for $R = 500$ and  (d)  Using surrogate data for $R = 28$.
   }
\label{svdlor}
\end{figure}

One of the difficulties of this technique is the choice of the
appropriate delay time, $\tau$, which in principle depends on
the specific nature of the system. 
To avoid this ambiguity,
a modified technique has been proposed by \cite{Bro86} where, 
a matrix constructed 
of vectors with unit delay time, is decomposed into eigenvectors
which then represent the dynamics. Application of this Singular Value Decomposition (SVD) technique
to the Lorenz system is shown in the top two panels of Figure \ref{svdlor}
which is again qualitatively similar to Figure \ref{lormap}.
To generate Figure \ref{svdlor}, the time series, $X(t)$ has been
converted to a uniform deviate, which is convenient when
the analysis is undertaken on data obtained from natural systems
(see next section). 

To confirm that the dynamical picture obtained by the Delay
Embedding technique is indeed due to an underlying deterministic non-linear
system and not of stochastic origin, it is customary to compare
the results obtained with those of surrogate data \citep{The92}. 
Here surrogate data are generated which have practically the
same power spectrum and distribution as the original time series, but
are of stochastic origin. 
The essential idea  is to formulate a null hypothesis 
that the data has been created by a stationary linear stochastic process, and 
then to attempt to reject this hypothesis by comparing results for the 
data with appropriate realizations of surrogate data. 
\cite{Sch96} have proposed a scheme, known as 
Iterative Amplitude Adjusted Fourier Transform (IAAFT), which generates
surrogate data that are more consistent in representing 
the null hypothesis \citep{Kug99}. In this work, the software which uses
tha above technique to generate surrogates 
(http://www.mpipks-dresden.mpg.de/tisean/TISEAN\_2.1/index.html)
has been used.
The bottom panel of Figure \ref{svdlor},
shows the SVD decomposition for surrogate data of the Lorenz system.
As expected, the dynamical phase picture is qualitatively different
for the surrogate as compared to the original time series. It should
be emphasized that a qualitative or quantitative difference between
a particular kind of surrogate and original data, although highly
suggestive, does not necessarily
imply that the original time series has a non stochastic origin. 
Ideally, the null hypothesis should be tested using 
several different kinds of surrogate data, before a concrete conclusion
can be reached.

\begin{figure}
\begin{center}
\includegraphics*[width=8cm]{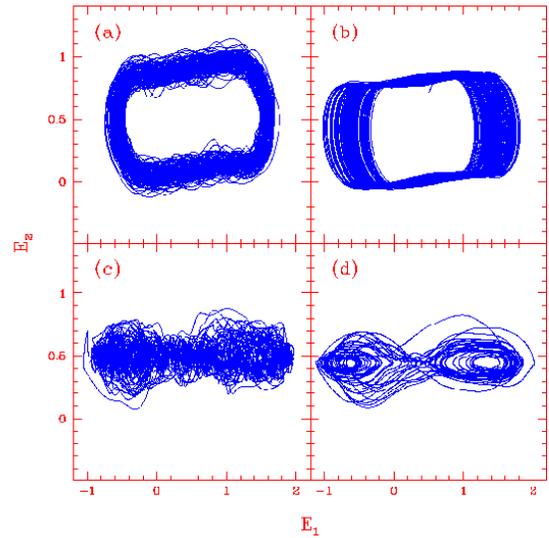}
\end{center}
\caption{ Singular Value Decomposition (SVD) technique
used to reconstruct the Lorenz system dynamics from $X (t)$ with
 noise contamination at 20\% level. 
(a) Using data for $R = 500$ with white noise, (b)  using data for $R = 500$
with red noise,
(c) using data for $R = 28$ with white noise and 
(d)  using  data for $R = 28$ with red noise.
  }
\label{svd_noise}
\end{figure}

Data from a
natural deterministic non-linear system may have contamination from
stochastic noise, either inherently or due to the detection process.
Such contamination generally makes it difficult to identify the
non-linear behavior of the system. Apart from the level
of contamination, the effect on the analysis also depends
on the nature of the noise. At the same level of contamination
(i.e. at the same percentage of noise addition)
the effect of white noise ( i.e. when the power spectrum of the
noise signal is a constant) is more than that of red
noise ( i.e. when the power spectrum, $P(f) \propto f^{-2}$).  This
is illustrated in Figure \ref{svd_noise}, where SVD plots of the
Lorenz system with 20\% red and white noise addition are shown. 
While the effect of noise on the limit cycle is to broaden
the original one dimensional curve, its effect on the  complex 
chaotic behavior can be more pronounced. Note that a 20\% white noise
contamination does not allow the dynamics to be 
effectively reconstructed for the Lorenz system with $R = 28$.

It is often desirable to have a quantitative measure apart from 
a qualitative picture of the reconstructed dynamics. Such a 
quantitative measure is the computation of the correlation dimension.
Briefly, the computation procedure is to choose a large number of
points in the reconstructed dynamics as centers. 
The correlation function is the number of points  which are within a
distance R from a center, averaged over all the centers, and may
be formally written as
\begin{equation}
C_M(R) \equiv \lim_{N \rightarrow \infty} \lim_{N_c \rightarrow \infty} {1\over N(N_c-1)} \sum_{i}^{N} \sum_{j, j \neq i }^{N_c}\hbox {H} (R-|\vec{x}_i -\vec{x}_j|)
\end{equation}
where $\vec{x}_j$ are the reconstructed vectors (Eqn 2), $N$ is the number 
of vectors, $N_c$ is the number of centers and H 
is the Heaviside step function. 
The fractional or correlation dimension $D_2 (M)$ is defined as 
\begin{equation}
  	D_2 (M) \equiv \lim_{R \rightarrow 0} \frac{d\hbox {log} C_M (R)}{d\hbox {log} (R)}  	
\end{equation}
and is essentially the scaling index of the variation of $C_M(R)$ with $R$.
With these definitions, for a finite duration light curve, 
only an approximate value of $D_2 (M)$ can be computed. Moreover
for small values of $R$, $C_M(R)$ is of order unity and hence will
not show the intrinsic scaling. Similarly, for large values of
$R$, $C_M(R)$ will saturate to the total number of data points.
Thus for a finite data set, an appropriate scaling region in $R$
needs to be chosen which does not suffer from the above edge effects, 
to obtain an approximate value of $D_2(M)$.
Here we use the non-subjective algorithm proposed by 
\cite{Mis04} which
establishes a region in $R$ which suffers least from edge effect errors 
and computes $D_2 (M)$ with an error estimation. A numerical code
which implements this algorithm is available at
http://www.iucaa.ernet.in/$^{\sim}$rmisra/NLD.

\begin{figure}
\begin{center}
\includegraphics*[width=8cm]{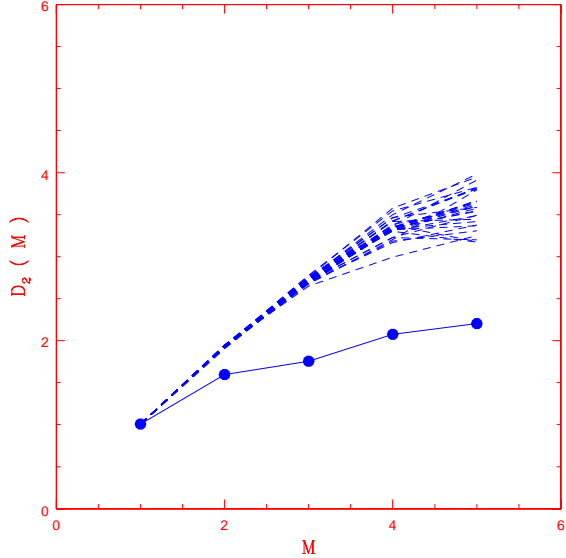}
\end{center}
\caption{ $D_2 (M)$ values (filled circles) for the Lorenz system exhibiting chaotic
behavior ( $R = 28$). The dashed lines represent $D_2 (M)$ curves
for thirty realizations of  surrogate data. The $D_2 (M)$ saturates to
$D_2^{sat} \approx 2$ and is different from the curves for surrogate data.  }
\label{lor_D2}
\end{figure}

\begin{figure}
\begin{center}
\includegraphics*[width=8cm]{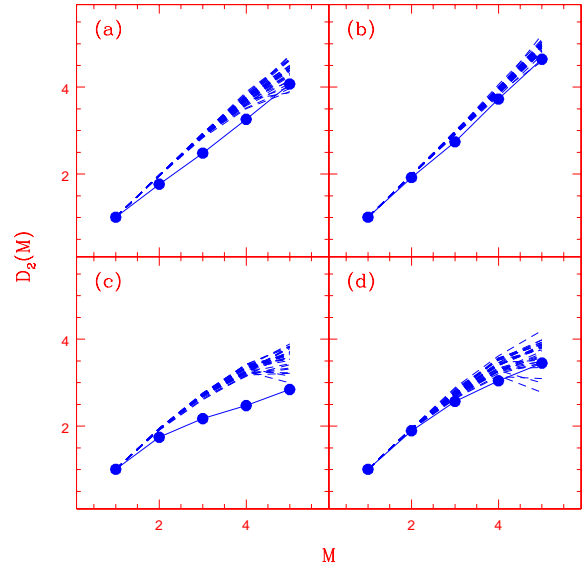}
\end{center}
\caption{ $D_2 (M)$ values for the Lorenz system and corresponding
surrogate with different levels of white and red noise contamination.
(a) 20\% white noise, (b) 50\% white noise, (c) 20\% red noise and (d) 50\% red noise.   }
\label{lor_D2_noise}
\end{figure}

For a chaotic
system, $D_2 (M) \approx$ 
constant $ = D_2^{sat}$ for $M$ greater than a certain dimension $M_{cr}$. 
For the Lorenz system exhibiting chaotic behavior (for e.g. when $R = 28$),
\cite{Spr01} have computed using a large number ($N > 100000$) points that
$D_2^{sat} = 2.05\pm 0.1$ and $M_{cr} = 3$. 
Figure \ref{lor_D2} shows the $D_2 (M)$ values for the Lorenz system
(for $R = 28$  using $5000$ points). As expected $D_2 (M)$ saturates to
$D_2^{sat} \approx 2$ for $M > M_{cr} \approx 3$. 
 For pure uncorrelated stochastic white noise
$D_2 (M) = M$ for all $M$ and no such saturation exists. However, a
saturated value  of $D_2 (M)$ does not necessarily imply that the
time series is due to deterministic non-linear behavior. In fact,  colored
noise (for which the power spectrum $P (f) \propto f^{-\alpha}$) 
exhibits saturation which can be computed to be $D_2^{sat} = 2/(\alpha-1)$ \citep{Osb89}.
Thus it is necessary to compare $D_2 (M)$ obtained from the time series
with corresponding surrogate data. This is done in Figure \ref{lor_D2},
where as expected the $D_2 (M)$ for the Lorenz system is significantly
different from the values for the surrogate data. For a deterministic non-linear 
system the computed $D_2^{sat}$ is a quantitative measure of the dynamics
and the critical dimension $M_{cr}$ is a measure of the number of
equations ( for the Lorenz system $M_{cr} = 3$) that are required to
describe the behavior.  Contamination by noise typically increases the
computed $D_2 (M)$ values and makes it more difficult to distinguish the
time series behavior from surrogate data. This is demonstrated in 
Figure \ref{lor_D2_noise}, where $D_2 (M)$ values are plotted for
the Lorenz system with different kinds and level of noise contamination.

\section{The Black Hole system GRS 1915+105}

\begin{figure}
\begin{center}
\includegraphics*[width=8cm]{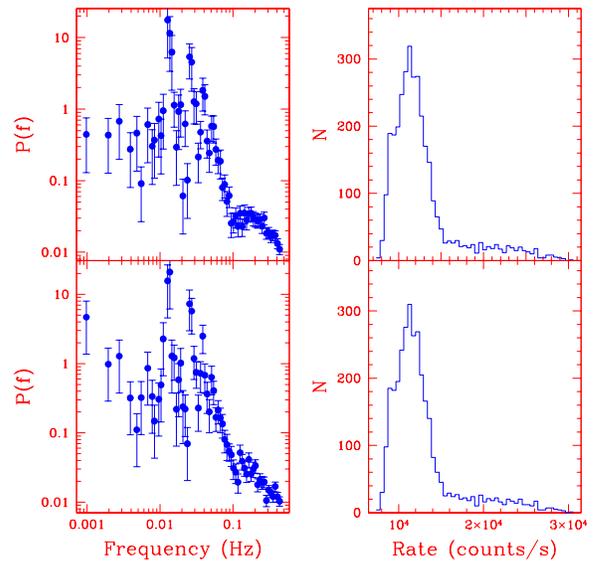}
\end{center}
\caption{ Top: The power spectrum and the flux ( count rate) distribution
for the X-ray light curve of GRS 1915+105 in the $\rho$ state. Bottom:
The power spectrum and the flux ( count rate) distribution
for corresponding surrogate data.   }
\label{GRS_rho_pow}
\end{figure}

Based on RXTE observations, \citet{Bel01} classified the various temporal
behavior of GRS 1915+105 into twelve classes. Here we take a representative
data for each class and generate continuous light curves in the energy range
$0-30$ keV binned at $1$ sec intervals. The details of the specific data used
in the following analysis are tabulated in  \cite{Mis04}. 
Figure \ref{GRS_rho_pow} shows the 
power spectrum of the X-ray light curve of 
GRS 1915+105 when the source is in the
$\rho$ state. The
presence of a low frequency oscillation 
with several harmonics is revealed. The flux (or the count rate) 
distribution during this state has a non Gaussian shape.  Although the
presence of harmonics of the oscillation and the flux distribution, 
suggests that the system is  non-linear \citep[e.g.][]{Utt05}, they do not 
necessarily imply that this behavior is due to a deterministic
non-linear system. The bottom panel of Figure \ref{GRS_rho_pow}
shows the power spectrum and flux distribution for corresponding
surrogate data and these are practically indistinguishable from the real case.
Thus non-linear time series analysis as described in the previous section
is required to ascertain whether the temporal behavior is due to 
 deterministic non-linear behavior.

\begin{figure}
\begin{center}
\includegraphics*[width=8cm]{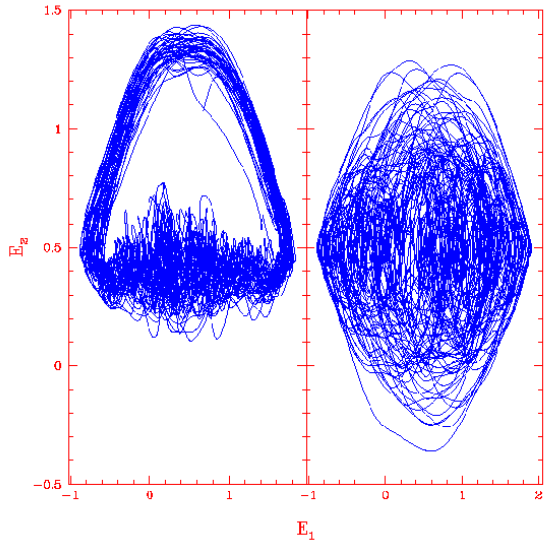}
\end{center}
\caption{ Singular Value Decomposition (SVD) technique
used to reconstruct the dynamics using the X-ray light curve
for GRS 1915+105 in the $\rho$ state. Left Panel is for real data
while the right panel is for a surrogate data. }
\label{GRS_rho_svd}
\end{figure}

The SVD reconstruction of the dynamics for
the data is shown in Figure \ref{GRS_rho_svd}. The presence of a main cyclic
feature is seen, which is similar to that of the Lorenz system with $R = 500$ (Figure \ref{svdlor}). 
This indicates that the observed oscillation is
of a limit cycle origin. Note that the surrogate data which has the
same power spectrum and distribution  has a different
SVD reconstruction (Figure \ref{GRS_rho_svd}).

\begin{figure}
\begin{center}
\includegraphics*[width=8cm]{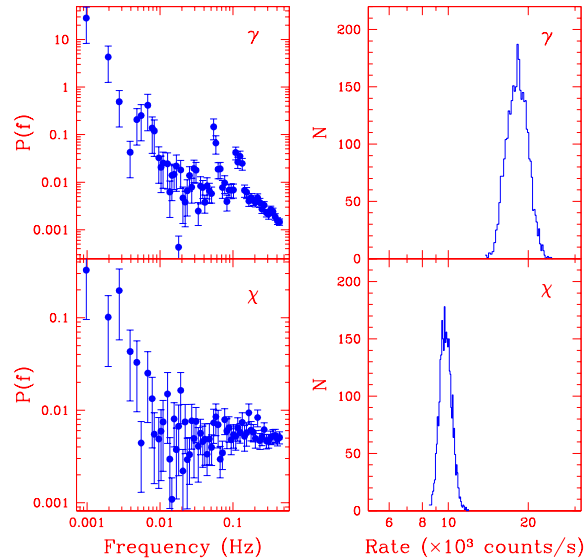}
\end{center}
\caption{ The power spectrum and the flux (count rate) distribution
for the X-ray light curve of GRS 1915+105 in the $\gamma$ and $\chi$ classes. 
}
\label{GRS_chi_pow}
\end{figure}

\begin{figure}
\begin{center}
\includegraphics*[width=8cm]{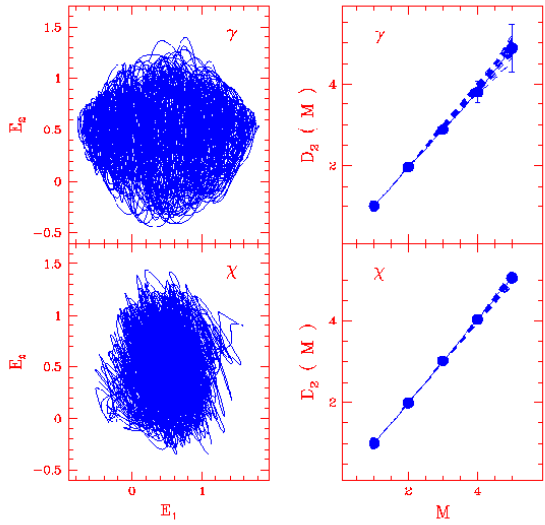}
\end{center}
\caption{ Left Panels: Singular Value Decomposition (SVD) technique
used to reconstruct the dynamics using the X-ray light curve
for GRS 1915+105 in the $\gamma$ and $\chi$ classes. Right panels:
$D_2 (M)$ values for the $\gamma$ and $\chi$ classes compared with
corresponding surrogates.
}
\label{GRS_chi_svd}
\end{figure}

The power spectra and flux distribution for two other class
of GRS 1915+105, $\gamma$ and $\chi$, are shown in Figure \ref{GRS_chi_pow}.
A low frequency QPO with harmonics similar to that seen
in the $\rho$ class is detected for the $\gamma$ class. During the
$\chi$ class no such QPO is detected.
However, for both these classes, the
SVD reconstruction of the dynamics does not show any qualitative 
features (Figure \ref{GRS_chi_svd}). Such space filling SVD reconstructions
are characteristics of systems with uncorrelated stochastic noise. In agreement with this, 
the $D_2 (M)$ values are $\approx M$ and  
show no saturation, for these two states (Figure \ref{GRS_chi_svd}). 
Moreover, the $D_2(M)$ are consistent with those obtained from 
corresponding surrogate data.  Data from the $\phi$ and $\delta$ classes show
similar behavior and are not shown. These results imply
that the variability of these classes are of stochastic nature, in
contrast to that of the $\rho$ class. In particular,
the observed low frequency QPO in the $\rho$ and $\chi$ class have
a different origin.

\begin{figure}
\begin{center}
\includegraphics*[width=8cm]{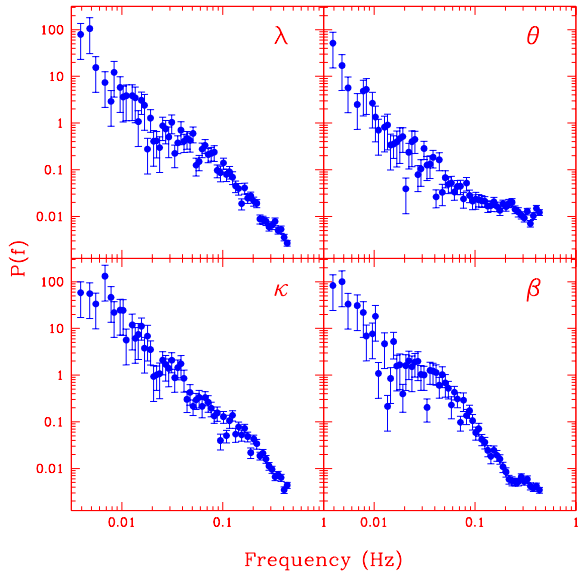}
\end{center}
\caption{ The power spectra 
for the X-ray light curve of GRS 1915+105 in the $\lambda$, $\theta$, $\kappa$ and $\beta$ classes. 
}
\label{GRS_beta_pow}
\end{figure}

\begin{figure}
\begin{center}
\includegraphics*[width=8cm]{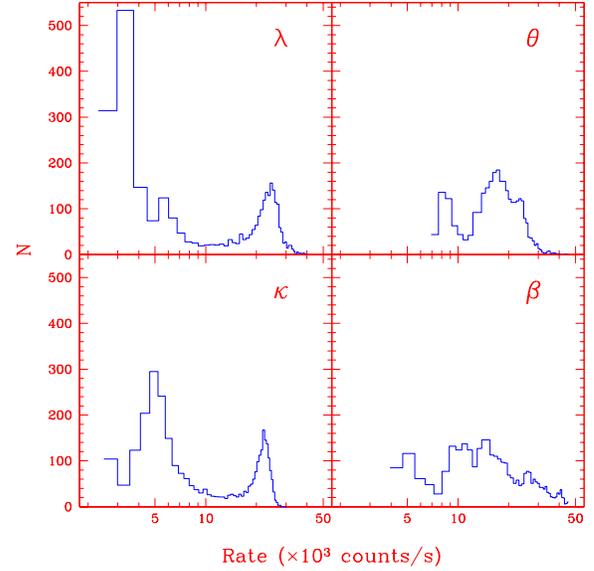}
\end{center}
\caption{ The flux (count rate) distribution
for the X-ray light curve of GRS 1915+105 in the $\lambda$, $\theta$, $\kappa$ and $\beta$ classes.
}
\label{GRS_beta_dist}
\end{figure}

\begin{figure}
\begin{center}
\includegraphics*[width=8cm]{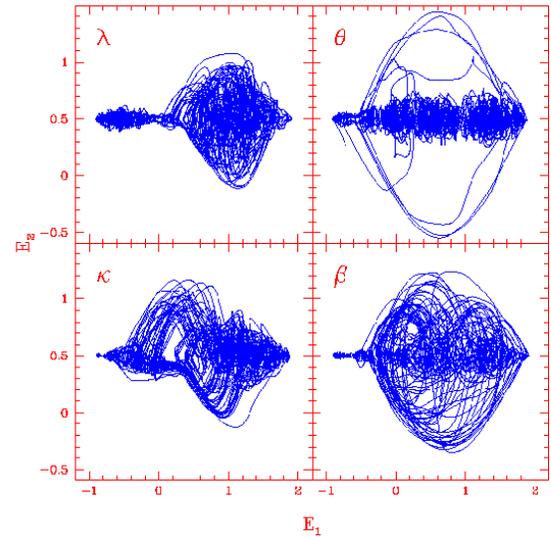}
\end{center}
\caption{Singular Value Decomposition (SVD) technique
used to reconstruct the dynamics using the X-ray light curve
for GRS 1915+105 in the $\lambda$, $\theta$, $\kappa$ and $\beta$ classes. 
}
\label{GRS_beta_svd}
\end{figure}

\begin{figure}
\begin{center}
\includegraphics*[width=8cm]{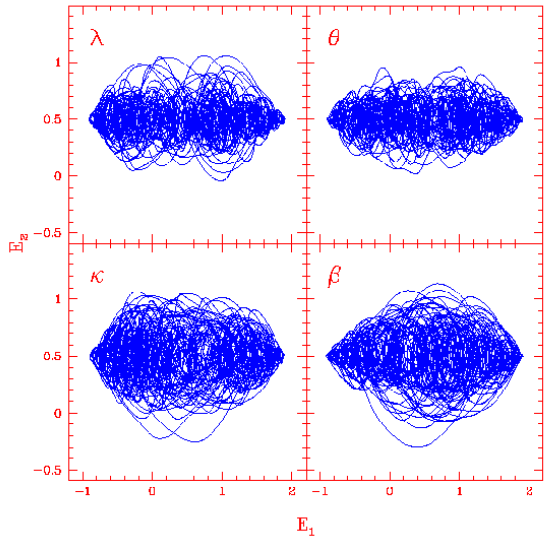}
\end{center}
\caption{ Singular Value Decomposition (SVD) technique
used to reconstruct the dynamics using surrogate data corresponding to 
the X-ray light curves
for GRS 1915+105 in the $\lambda$, $\theta$, $\kappa$ and $\beta$ classes. 
}
\label{GRS_beta_svd_surgg}
\end{figure}

\begin{figure}
\begin{center}
\includegraphics*[width=8cm]{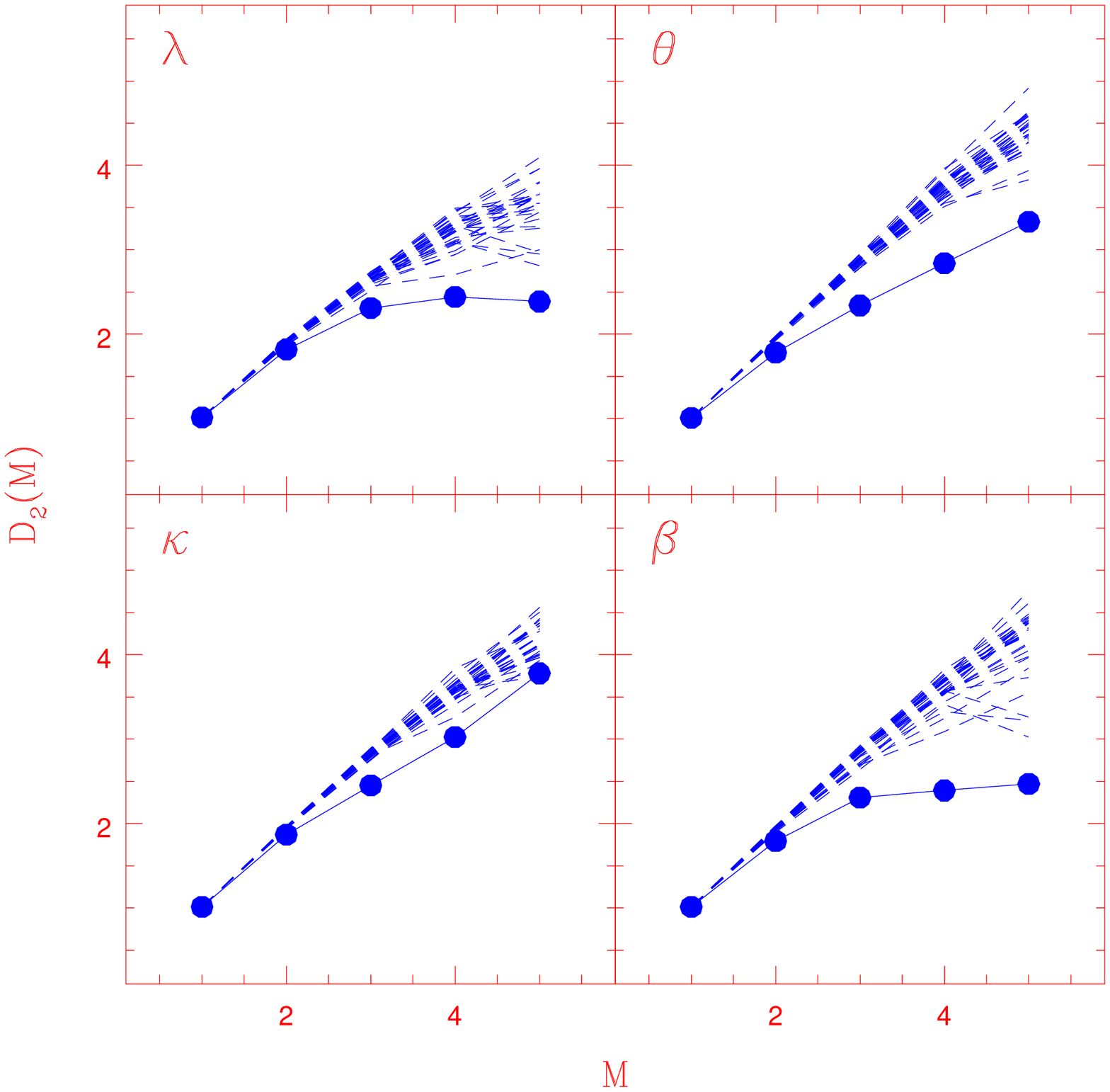}
\end{center}
\caption{ 
$D_2 (M)$ values for the $\lambda$, $\theta$, $\kappa$ and $\beta$ classes compared with
corresponding surrogates.
}
\label{GRS_beta_d2}
\end{figure}

The behavior of the rest of the classes is more complex. For example,
the power spectra for $\lambda$, $\theta$, $\kappa$ and $\beta$ class
do not show any strong low frequency periodicity (Figure \ref{GRS_beta_pow}),
while the flux distribution is typically non-Gaussian 
(Figure \ref{GRS_beta_dist}). The SVD reconstruction of these data (Figure \ref{GRS_beta_svd}) show
structure which qualitatively maybe similar to that of a chaotic
system with noise (Figure \ref{svd_noise}) and are somewhat different than
the SVD reconstruction of the corresponding surrogate data 
(Figure \ref{GRS_beta_svd_surgg}). More evidence is provided by 
the $D_2 (M)$ values for these data sets (Figure \ref{GRS_beta_d2})
which show saturation and are different from the values computed from
corresponding surrogate data.

\section{Summary and Discussion}

Using non-linear time series analysis techniques, evidence has been
provided that the black hole system GRS 1915+105 exhibits stochastic
variability when it is in $\chi$, $\gamma$, $\phi$ and $\delta$ classes.
In the $\rho$ class, the low frequency QPO detected in the power spectrum
can be attributed to a non-linear limit cycle behavior. For the
other classes, the system shows behavior similar to a non-linear chaotic
system with some inherent stochastic noise.

Before considering the implications, it is prudent to emphasize that 
the results presented here need to be confirmed by more
comprehensive non-linear data analysis. Since the results are based
on a null hypothesis test involving surrogate data, a more comprehensive
test should take into account possible different kinds of surrogate
data.

The saturation of $D_2 (M)$ to $D_2^{sat} \approx 3-4$ (Figure 
\ref{GRS_beta_d2}) in analogy with that of the Lorenz system
with noise (Figure \ref{lor_D2_noise}), implies that the
temporal behavior of the system is governed by 3-4 underlying 
global equations. On the other hand, it is believed that the temporal
evolution of an accretion disk can be represented by 
local magneto-hydrodynamic equations which are non-linear 
differential equations in both time and disk radius. Thus,
an encouraging aspect of the results obtained here is that these
complicated magneto-hydrodynamic equations may be 
reducible to (or at least approximated by) a set of simple (although
still non-linear) global equations, which are functions of
global parameters like size of the disk, average density, pressure etc.
Such a set of global equations, will provide a complete picture of
the global properties of the inner accretion disk and in principle may enable 
the testing of general relativity in the strong field limit. However,
obtaining these equations from the magneto-hydrodynamic equations
directly is challenging. Perhaps spatial averaging of these equations
to obtain a one (or two zone) models like as done by \citet{Pac83} may be
the most promising option. Such approximations may be obtained with 
insights gained  from the results of future 
sophisticated numerical magneto-hydrodynamic simulations of realistic
accretion disks.

Non-linear time series analysis can also be used to compare results
of numerical simulations with real data. If the simulation correctly
encompasses the basic non-linear behavior of the system, the resultant
simulated light curve should have similar $D_2 (M)$ values and SVD
reconstruction. 

In analogy with the Lorenz system, the different temporal behavior
(corresponding to the different classes) can be attributed to
variation of a single control parameter, which for the Lorenz system
is $R$ (Eqn \ref{loreqn}). For GRS 1915+105, this controlling parameter
could be the time averaged accretion rate, which is determined by
the outer boundary condition. One of the key attributes of a 
deterministic non-linear system is that the time variability need
not be due to a time varying parameter. For example, the equations 
defining the Lorenz system (Eqn \ref{loreqn}) do not have an explicit
time dependent term. For GRS 1915+105, it is possible that the different
temporal classes arise from variations in the accretion rate, while
the variability for a given class is due to an inherent non-linearity.
Thus it is not necessary to postulate other varying parameters, apart
from the accretion rate (like magnetic field etc) to explain the source's
complex variability. If this is true, then the average bolometric luminosity
which should be proportional to the average accretion rate, should be 
different for each class. However, the bolometric luminosity can be
estimated only by model dependent spectral fitting with the 
caveat that the time-averaged spectra may not be adequately represented
by steady ones. Nevertheless, the results presented here provide the
possibility of identifying a single control parameter 
which determines the temporal class exhibited by the system.

The temporal behavior of GRS 1915+105 on time-scales of hours
and months is unique, but
may be related to the state changes that occur for other black hole systems
(like Cygnus X-1) on time-scales of months. This difference between
GRS 1915+105 and the other black holes has led to the speculation that
GRS 1915+105 has a rapidly spinning black hole.
Identifying the governing
non-linear equations for GRS 1915+105 will not only provide insight
into the temporal variability of other black hole systems, but will
perhaps give an explanation for its uniqueness.

\end{document}